\documentclass[twocolumn,showpacs]{revtex4}

\usepackage{graphicx}
\usepackage{dcolumn}
\usepackage{amsmath}
\usepackage{epsfig}
\usepackage{amssymb}
\usepackage{theorem}

\usepackage{hhline}

\theoremstyle{plain}

\theoremstyle{plain}

\numberwithin{equation}{section}

\newcommand{\ket}[1]{\left\vert#1\right\rangle}
\newcommand{\bra}[1]{\left\langle#1\right\vert}

\newcommand{\beq}{\begin{equation}}
\newcommand{\eeq}{\end{equation}}
\newcommand{\bea}{\begin{eqnarray}}
\newcommand{\eea}{\end{eqnarray}}

\makeatletter
\def\btt#1{\texttt{\@backslashchar#1}}
\DeclareRobustCommand\bblash{\btt{\@backslashchar}}
\makeatother

\begin{document}

\title{High Fidelity State Transfer Over 
an Unmodulated Linear $XY$ Spin Chain}

\author{C. Allen Bishop$^{1}$\footnote{Electronic address: abishop@www.physics.siu.edu}, Yong-Cheng Ou$^{1}$, 
Zhao-Ming Wang$^{2}$, and Mark S. Byrd$^{1}$}  
\affiliation{$^{1}$ Physics Department, Southern Illinois University, 
Carbondale, Illinois 62901-4401}
\affiliation{$^{2}$ Physics Department, Ocean University of China, Qingdao, 
266100, China}

%\email{abishop@www.physics.siu.edu}

\date{\today}

\begin{abstract}
We provide a class of initial encodings that can be sent with a high 
fidelity over an unmodulated, linear, $XY$ spin chain. As an example, 
an average fidelity of ninety-six percent can be obtained using 
an eleven-spin encoding to transmit a state over a chain containing 
ten-thousand spins. An analysis of the magnetic field dependence 
is given, and conditions for field optimization 
are provided.   
\end{abstract}


\maketitle

%-------------------------------------------------------------------------
%-------------------------------------------------------------------------

\section{Introduction}

The notion of using an unmodulated spin chain to serve as a channel 
for the transmission of quantum information was put forth by 
Bose in 2003 \cite{Bose:03}. 
To improve the communication fidelity of the original proposal, 
a considerable effort has been made in order to find ways of preserving  
the integrity of a quantum state as it 
propagates across these quantum wires. It was recently shown that, 
in principle, perfect 
state transfer can be obtained between processors that are connected 
by any interacting media \cite{Wu/etal:09}. Specific protocols which yield 
ideal state transmissions have also been proposed using quantum dots 
\cite{Petrosyan:04a,Petrosyan:04b} and spin-chain systems with
non-identical couplings between pairs 
\cite{Christandl/etal:04,Christandl/etal:05}. It has 
also been shown that the fidelity can be increased if one can 
dynamically control the first and last pairs of the chain 
\cite{Haselgrove:05}. In fact, the fidelity can become arbitrarily 
large using this last method if there is no limit to the 
number of sequential gates that can be applied \cite{Burgarth:07}.
Perfect state transfer can also be obtained without prior initialization 
of the spin medium if one applies end gates to a chain 
constructed with pre-engineered couplings \cite{Franco/etal:08}. 
Furthermore, the communication can also be 
improved if the sender encodes the message 
state over the space of multiple spins rather than only one 
\cite{Osborne/etal:04,Haselgrove:05,Allcock/etal:09,Wang/etal:09}. 
(For a nice review of the subject see Ref.~\cite{Bose:07}.)

In Ref.~\cite{Osborne/etal:04}, 
the authors cleverly chose to use the all-spin-down eigenstate to 
represent the $\ket{0}$ basis state of an encoded qubit and 
were able to find single excitation encodings for $\ket{1}$
that propagated 
with little dispersion through a spin ring. In their 
proposal both the sender and receiver have access to multiple 
spins.
They also assumed that the receiver has 
the ability to unitarily focus the entire probability 
amplitude of finding the excitation within his region 
onto a single site. This method can enhance the transfer fidelity to values 
well above those which can be obtained using the original single 
spin encoding scheme \cite{Bose:03}.

Using this protocol Haselgrove 
derived a method to obtain 
the initial encodings which maximize the fidelity of state 
transmission \cite{Haselgrove:05}. His method was based on the singular-value 
decomposition and can be applied to systems that 
conserve the total z-component of spin. In 
particular, his method can be applied to a chain of spin-1/2 particles 
coupled via the $XY$ 
interaction. His encodings are optimal in the sense that they 
maximize the probability amplitude of finding the transmitted 
excitation somewhere within the receivers accessible region. This 
is a sufficient condition given the assumption that the receiver 
can perform the necessary decoding operation mentioned above. In 
this protocol the fidelity is measured with respect to a 
single qubit output state although the initial encoding has been 
carried out using multiple spins.

In Ref.~\cite{Wang/etal:09}, a particular state was found to transfer 
well across relatively long $XY$ coupled spin chains. It was shown 
that if the first and third spins were placed in 
the singlet state a fidelity of ninety percent or better could 
be obtained for chains consisting of up to fifty spins. The 
high fidelities do not depend on the dynamical 
control of the chain nor do they require the prefabrication of 
special couplings for each neighboring pair. The fidelities 
were determined according to the direct overlap between the received 
state and the actual state which was sent, i.e., the receiver 
was not required to implement a decoding 
unitary. Given the assumption that the chain was placed in the all-spin-down 
ground state prior to initialization, one can reliably transmit 
information to the receiving end using a simple two spin encoding.

Motivated by this result, we have found a class of effective 
$k$-qubit encodings ($k = 2,3,\hdots$) which can be used to 
reliably send an encoded qubit state over very long chains. Each 
member of this class has a structure similar to the singlet 
encoding of \cite{Wang/etal:09}, with each increase in $k$ yielding 
higher fidelities. We take these states to represent the $\ket{1}$ 
basis of a logical qubit, with $\ket{0}$ taken to be the all-spin-down 
eigenstate. The paper will focus on the least technically challenging $XY$ 
configuration; we consider linear arrays of spins having 
constant and equal exchange couplings between neighboring pairs. 
As in \cite{Wang/etal:09}, we will not require the receiver to 
implement a decoding operation, we simply let the encoded states 
propagate freely across the chain. As an example of the reliability 
of these states, we find that an eleven-qubit encoding can 
be sent across a chain containing ten-thousand spins and arrive 
with an average fidelity of ninety-six percent. We analyze the 
influence of an external magnetic field and find that by
isolating the system one can maximize the fidelity provided that the 
chain has an appropriate number of sites.

We also compare our results to those which can be obtained 
using Haselgrove's method. We find that for small chains his 
optimal states give slightly higher fidelities than ours when 
the encodings are viewed to take place over the first $r$ spins 
($r=3,4,\hdots$) of the chain.  However, our encoded states only use a
subset of the first $r$ spins, generally $(r+1)/2$ spins are used. 
If we compare our $n$-spin encoding to Haselgrove's $n$-spin encoding,
our method yields higher fidelities.  
Again, we emphasize that his fidelities are 
determined with the assumption that the receiver has applied 
a decoding operation while for our states no such requirement 
is necessary. For long chains our encodings actually converge 
to Haselgrove's optimal encodings and it is found that the fidelities are 
equal before and after the decoding process has taken place. 

The second section of this paper will provide the necessary background 
for what follows. In Sec. III we will introduce this class of states 
and provide examples of the high fidelities which can be obtained 
when transferring members of this class over long chains. Sec. IV 
extends the analysis to encoded qubit states and discusses 
the associated average fidelities. That section also includes 
a discussion of the dependence of the fidelity on a global 
magnetic field. In Sec. V we will compare our results to those 
using Haselgrove's method. Finally, we will conclude with a summary 
of our results in Sec. VI.

%---------------------------------------------------------------
%---------------------------------------------------------------

\section{Unitary Evolution of a Spin Chain: XY Model}

We will consider a linear chain of qubits with uniform, 
time-independent nearest neighbor 
couplings. The chain is subjected to a global magnetic field 
$h$ (aligned along the $z$-axis) which is assumed to remain static 
over time. The Hamiltonian function of this $N$-spin system is
\beq
\label{eq:Ham}
H = -\frac{J}{2}\sum_{i=1}^{N-1} (\sigma_i^x \sigma_{i+1}^x + 
\sigma_i^y \sigma_{i+1}^y) -h\sum_{i=1}^N \sigma_i^z,
\eeq
where $\sigma_i^x$ and $\sigma_i^y$ are pauli matrices acting on the 
$i$th spin and $J$ determines the coupling strength. 

Following the usual convention, we will let $\ket{{\bf{j}}}$ 
$({\bf{j}} = {\bf{1,2,...,N}})$ represent the state where all of the 
spins in the chain point ``down'' except for the spin at site 
${\bf{j}}$ which has been flipped to the ``up'' state. The 
ground state will be denoted as $\ket{{\bf{0}}}= 
\ket{\downarrow \downarrow \downarrow ...\downarrow \downarrow \downarrow}$. 
In this paper we will focus strictly on the evolution of states 
encoded in the Hilbert spaces ${\cal{H}}^{(0)}$ and ${\cal{H}}^{(1)}$, where 
${\cal{H}}^{(0)}$ (${\cal{H}}^{(1)}$) are respectively spanned by the states 
$\{\ket{{\bf{0}}} \}$ ($\{\ket{{\bf{j}}}: 
{\bf{j}} = {\bf{1,2,...,N}} \}$). Since the Hamiltonian $H$ 
(Eq.~(\ref{eq:Ham})) commutes with $\sum_{i=1}^N \sigma_z^i$, states 
which are encoded into either space ${\cal{H}}^{(0)}$ or ${\cal{H}}^{(1)}$ 
will remain there in the absence of noise.

An arbitrary initial encoding over the first $r$ spins of the chain can 
be expressed in the notation above as 
\beq
\label{eq:in}
\ket{{\bf{\Psi}}(0)} = \sum_{j=0}^{r} 
\alpha_j \ket{{\bf{j}}},
\eeq 
with $\sum_{j=0}^{r} |\alpha_j|^2  = 1.$
This state evolves unitarily to   
\beq
\ket{{\bf{\Psi}}(t)} = \alpha_0 \ket{{\bf{0}}} +
\sum_{j=1}^{N} w_j(t) \ket{{\bf{j}}},
\eeq  
where $w_j(t) = \sum_{s=1}^{r} 
\alpha_s \bra{{\bf{j}}}e^{-iHt}\ket{{\bf{s}}} = 
\sum_{s=1}^{r} \alpha_s f_{s,j}(t).$ (We let $\hbar = 1$ throughout.) 
The transition amplitudes 
$f_{s,j}(t)$ are given by
\beq
\label{eq:ta}
f_{s,j}(t) = \frac{2}{N+1}\sum_{m=1}^N \sin{(q_m s)} \sin{(q_m j)} e^{-iE_m t},
\eeq
with $E_m = 2h -2J\cos{(q_m)}$, and $q_m = \pi m/(N+1)$. 

Given the 
coefficients $\alpha_j$ of an initial encoding along 
with the transition amplitudes provided in Eq.~(\ref{eq:ta}) one 
can calculate the fidelity of state transfer through an unmodulated, 
linear, XY 
spin chain. We will provide an expression for the fidelity in the next 
section and introduce a class of states which travel exceptionally 
well over a chain composed of a large number of sites.

%---------------------------------------------------------------
%---------------------------------------------------------------

\section{High Fidelity Transfer of a Class of States }

It was recently shown in Ref.~\cite{Wang/etal:09} that an
initial encoding of $\ket{{\bf{\Psi}}(0)} = 
(\ket{{\bf{1}}} - \ket{{\bf{3}}}) /\sqrt{2}$ can be 
transferred with a relatively high fidelity $(F \approx .9)$ to the 
opposite end of an 
unmodulated spin chain containing $N \approx 50$ sites. Since we 
are assuming that prior to encoding the entire chain has 
been cooled to its ground state, initializing $\ket{{\bf{\Psi}}(0)} = 
(\ket{{\bf{1}}} - \ket{{\bf{3}}}) /\sqrt{2} = 
\frac{\:1}{\sqrt{2}}(\ket{\downarrow \uparrow}_{1,3} 
-\ket{\uparrow \downarrow }_{1,3}) 
\otimes \ket{\downarrow \downarrow \downarrow ...  
\downarrow \downarrow \downarrow  }_{2,4,5,...,N}$ is effectively 
a two qubit process. This encoding has a very simple structure; 
with the exception of the singlet placed at the first and third site, 
every qubit remains in the ground state. 

This state belongs to a class of states which consists of effective 
$k$-qubit encodings ($k=2,3,4,...$) each having a similar structure. 
We will label the state which corresponds to a 
specific $k$ as $\ket{\Psi_k}$ and write them explicitly as 
\beq
\label{eq:statek}
\ket{\Psi_k} = \frac{\;1}{\sqrt{k}}\sum_{m=0}^{k-1} (-1)^m \ket{{\bf{2m+1}}},
\;\;  k=2,3,4,...
\eeq
Each successive increase in $k$ yields a higher fidelity of transmission 
as the spin chain grows large. To show this let us first express 
each member of this class in the form $\ket{\Psi_k} = \ket{\phi_k} \otimes 
\ket{\downarrow \downarrow \downarrow ... \downarrow \downarrow \downarrow}$,
where $\ket{\phi_k}$ describes the state of the first $r = 2k-1$ spins 
of the chain. Ideally, at some later time the chain would evolve to 
$\ket{\downarrow \downarrow \downarrow ... \downarrow \downarrow \downarrow} 
\otimes \ket{\phi_k}$ in which case a perfect transmission would result. 
The fidelity between the ``encoded'' state $\ket{\phi_k}$ 
and the state corresponding to the last $r$ spins of the 
chain $\rho(t)$ is given by $F=\sqrt{\bra{\phi_k}\rho(t)\ket{\phi_k}}$. 
Here $\rho(t)$ is the reduced density matrix associated with the state 
of the receiver's spins and is obtained by tracing over all but the 
last $r$ sites of the chain. Since a perfect transmission is described by
$\ket{\downarrow \downarrow \downarrow ... \downarrow 
\downarrow \downarrow} 
\otimes \ket{\phi_k}$, the state of the first spin of the encoding would 
ideally propagate to the $[N-(r-1)]{\it{th}}$ site of the chain while the 
state of the last spin of the encoding (the spin at the $r {\it{th}}$ 
site) would 
ideally propagate to the $N{\it{th}}$ site of the chain. For 
example, when $k=2$ the initial encoding given by 
$(\ket{{\bf{1}}} - \ket{{\bf{3}}})/\sqrt{2} = (\ket{\uparrow \downarrow \downarrow} - 
\ket{\downarrow \downarrow \uparrow})/\sqrt{2} \otimes 
\ket{\downarrow \downarrow \downarrow ... \downarrow \downarrow \downarrow}
$ would ideally evolve to the state $
\ket{\downarrow \downarrow \downarrow ... \downarrow \downarrow \downarrow} 
\otimes (\ket{\uparrow \downarrow \downarrow} - 
\ket{\downarrow \downarrow \uparrow})/\sqrt{2} = 
(\ket{{\bf{N-2}}} - \ket{{\bf{N}}})/\sqrt{2}$. 

In general, the fidelity between an initial encoding given by 
Eq.~(\ref{eq:in}) and the received state can be expressed as
\beq
F = \sqrt{|G|^2 + |\alpha_0|^2 \sum_{i=1}^{N-r}|w_i|^2}
\eeq
where $G = |\alpha_0|^2 + \sum_{i=1}^r \alpha_i w_{(N-r+i)}^*$, 
and $w_j = \sum_{i=1}^{r} \alpha_i f_{i,j}(t)$.
In this case we have 
$\alpha_{\nu} = 0$ for even $\nu$ and $\alpha_{\nu} = \pm 1/\sqrt{k}$ for 
odd $\nu$. 

Table I  lists the maximum fidelity achievable $F$ within the 
time interval $[0,N]$ along with 
the associated arrival times $t_0$ for 
the encodings $\ket{\Psi_k}$ ($k=2,3,4,5$). The table shows that 
as the number of spins grows large the 
fidelity increases as the value of $k$ increases. We can also 
see that an increase in the number of spins does not necessarily 
imply a decrease in the fidelity. In fact, increasing $N$ can actually 
increase $F$ on occasion, e.g., notice this increase for the 
states $\ket{\Psi_4}$ and $\ket{\Psi_5}$. Also, for a given 
number of spins the time in which this maximum fidelity is obtained 
is nearly equal for neighboring values of $k$. 

%The arrival time is slightly 
%shorter for $\ket{\Psi_{k+1}}$ compared to $\ket{\Psi_k}$ 
%due to the fact that the former state is encoded over more spins, and is
%thus initially closer to the opposite end. 

\begin{table}[tbp]
\label{tab:f}

\begin{tabular}{ccccccccccccccccccccccccccccccc}   
 && && &&  $k=2$  && && &&

\end{tabular} 

\begin{tabular}{|c|cccccccccccc|cccccccccccccccccc}

\hline 
$N$    && 100  && 200  && 300   && 400   && 500   && 600   

\\ \hline
$F$     && .83 && .74 && .68  && .64  && .60  && .57   

\\ \hline
$t_0$ && 51.75 && 102.36 && 152.76 && 203.07 && 253.33 && 303.55 \\

\hline

\end{tabular} 

\begin{tabular}{ccccccc}   \\

 $k = 3$  \\  

\end{tabular} 

\begin{tabular}{|c|cccccccccccc|cccccccccccccccccc}  \hline

$N$    && 100  && 200  && 300   && 400   && 500   && 600   

\\ \hline
$F$    && .90 && .88 && .84  && .81  && .78  && .75  

\\ \hline
$t_0$ && 51.08 && 101.74 && 152.20 && 202.55 && 252.84 && 303.09 

\\ \hline 

\end{tabular} 

\begin{tabular}{ccccccccccccc|cccccccccccccccccc} \\

$k = 4$ 

\end{tabular} 

\begin{tabular}{|c|cccccccccccc|cccccccccccccccccc}  \hline

$N$    && 100  && 200  && 300   && 400   && 500   && 600   

\\ \hline
$F$    && .88 && .90 && .90  && .88  && .87  && .85  

\\ \hline
$t_0$ && 51.12 && 101.10 && 151.54 && 201.91 && 252.22 && 302.49 
\\ \hline

\end{tabular} 

\begin{tabular}{ccccccccccccc|cccccccccccccccccc}  \\

$k = 5$  

\end{tabular} 

\begin{tabular}{|c|cccccccccccc|cccccccccccccccccc}  \hline

$N$    && 100  && 200  && 300   && 400   && 500   && 600   

\\ \hline
$F$    && .93 && .88 && .89  && .90  && .90  && .89  

\\ \hline
$t_0$ && 51.50 && 101.22 && 151.02 && 201.27 && 251.56 && 301.83 

\\ \hline 
\end{tabular} 

\caption{Table showing the maximum fidelity achievable and the associated 
arrival times for the states $\ket{\Psi_k}$, $k=2,3,4,5$ as 
a function of $N$. Note that the time has been 
calculated using $J=1.$ }
\end{table}

\begin{figure}[tbp]
\includegraphics[width=.45\textwidth]{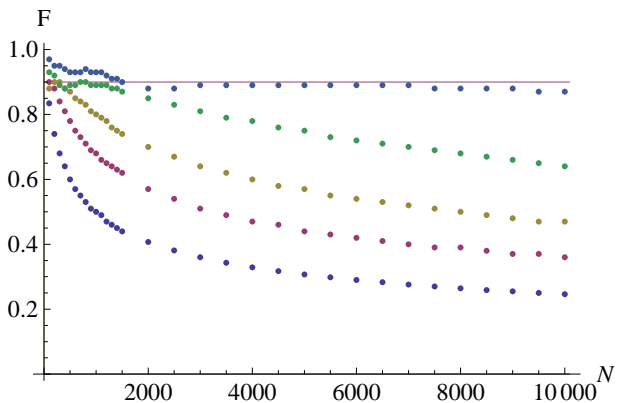}
\caption{(Color online) Maximum fidelity which can be obtained as 
a function of $N$ for the states $\Psi_2$,$\Psi_3$,
$\Psi_4$, $\Psi_6$, and $\Psi_{11}$. (From bottom to top.  
There are no crossings.) The horizontal line lies at 90$\%$. }
\label{fig:large}
\end{figure}

This trend of increasing fidelity with increasing $k$ continues 
as the number of spins gets very large. In Fig.~\ref{fig:large} 
we plot the maximum fidelity which can be obtained for several 
states in this class as a function of $N$. The curves shown there 
correspond, from bottom to top, to the states $\ket{\Psi_2}$, 
$\ket{\Psi_3}$, $\ket{\Psi_4}$, $\ket{\Psi_6}$, and $\ket{\Psi_{11}}$. 
For low values of $k$ ($k < 7$), an increase from $k$ to $k+1$ leads to 
an approximately $10\%$ increase in $F$. When $k$ gets larger than 6 
an increase from $k$ to $k+1$ still yields higher obtainable fidelities, 
although the rate of change steadily decreases. For example, 
when $N = 10,000$ the maximum fidelity which can be obtained 
when $k = 2,3,4,6,$ and $11$ is respectively $25\%$, $36\%$, $47\%$, 
$64\%$, and $87\%$. We may increase 
the value of $k$ beyond 11 to obtain even higher fidelities, 
the tradeoff, of course, comes at the expense of the additional 
resources needed for encoding.

Since we have taken both $\hbar$ and $J$ to be one the values 
associated with the times $t$ are simply unitless numbers. Specific 
values for the nearest neighbor exchange constants associated 
with several chemical compounds are listed in \cite{Hase/etal:04} 
and range in absolute value from one to several hundred Kelvin. In 
order to translate the tabulated values of $t_0$ into realistic 
values of time we consider a chain composed of the compound 
$({\text{N}}_2{\text{H}}_5){\text{CuCl}}_3$ which has $J=4.1\;K$. 
For $N=100$, the $k=2$ encoding 
$\ket{\Psi_2}$ yields a maximum fidelity of .83 at time $t_0 = 1.12 \mu s$ 
within the interval $t \in \left[0, \frac{N \hbar}{J}\right]$.

Since the ground state $\ket{{\bf{0}}}$ is an eigenstate of 
the Hamiltonian, and we have assumed that before the encoding 
process begins the chain has been cooled to this state, it makes 
sense to allow $\ket{{\bf{0}}}$ to represent one of the basis 
states of an encoded qubit. Let us define an encoded qubit 
which is spanned by the ground state and the $k$th member of this 
class to be
\beq
\label{eq:qbit}
\ket{\xi_k}= \cos{(\theta/2)}\ket{{\bf{0}}}
+e^{i\phi}\sin{(\theta/2)}\ket{\Psi_k}.
\eeq    
We will see in the next section how a global magnetic field can 
be used to increase the average fidelity of communication 
to values well above those which can be obtained using the states 
$\ket{\Psi_k}$ alone.

\vspace{.09in}

%-----------------------------------------------------------
%-----------------------------------------------------------

\section{Average Fidelity and the Magnetic Field}

So far we have not mentioned the magnetic field contribution 
to the Hamiltonian (Eq.~\ref{eq:Ham}). This is because 
the effect of such a field is to apply equal phase shifts 
to all initial encodings which lie in ${\cal{H}}^{(1)}$ and so 
the fidelities discussed thus far have been independent 
of $h$. We will now consider an encoded qubit which not only 
lies in ${\cal{H}}^{(1)}$ but in  ${\cal{H}}^{(0)}$ as well. 
${\cal{H}}^{(0)}$ and ${\cal{H}}^{(1)}$ are distinct eigenspaces 
of the operator of the total $z$ component of the spin; 
$\sigma_{tot}^z = \sum_{i=1}^N \sigma_i^z$. ${\cal{H}}^{(0)}$ is 
spanned by a single state, namely $\ket{{\bf{0}}}$, which is 
also an eigenstate of the total Hamiltonian $H$. The associated 
energy of the ground state is chosen to be zero and therefore the 
coefficient attached to $\ket{{\bf{0}}}$ in Eq.~(\ref{eq:qbit}) will not 
change over time. 

A global magnetic field is related to the fidelity through the 
transition amplitudes $f_{s,j}(t)$ in Eq.~(\ref{eq:ta}). For what 
follows let us re-express these amplitudes as 
$f_{s,j}(t) = e^{-2iht}\tilde{f}_{s,j}(t)$, where 
$\tilde{f}_{s,j}(t) = 
\frac{2}{N+1}\sum_{m=1}^N \sin{(q_m s)} \sin{(q_m j)} e^{2iJt\cos{(q_m)}}$ 
is independent of $h$. The fidelity $F_k$ of state transfer 
for the encoding $\ket{\xi_k}$ can be calculated to be 
\begin{widetext}
\bea
\label{eq:Fk}
F_k &=& \left[ \cos^4{\frac{\theta}{2}} + \frac{1}{2k}\sin^2{\theta}
 \left( \text{Re}
\left[e^{2iht}\sum_{m=1}^{k} (-1)^{m+1} C^*_{N+2(m-k)}(t) \right] 
+ \sum_{m=1}^{N+1-2k}|C_{m}(t)|^2
\right) \right. 
\nonumber \\
&+& \left. \frac{1}{k^2}\sin^4{\frac{\theta}{2}} 
\left| \sum_{m=1}^k (-1)^{m+1} C^*_{N+2(m-k)}(t) \right|^2 
\right]^{1/2},
\eea
\end{widetext}
where $C_{\nu}(t) = \sum_{p=0}^{k-1} (-1)^p \tilde{f}_{2p+1,\nu}(t).$ Since 
the $C_{\nu}(t)$ are independent of the applied field $h$ the 
dependence of $F_k$ on this quantity comes strictly from the 
second term in Eq.~(\ref{eq:Fk}). If we let 
$L(t) =\frac{1}{k}\sum_{m=1}^k (-1)^{m+1} C^*_{N+2(m-k)}(t)$ we can express 
the second term of Eq.~(\ref{eq:Fk}) as 
\beq
\label{eq:fhk}
F_{h,k} =
\frac{\sin^2{(\theta)}}{2}\left(\cos{(2ht)}\text{Re}\left[L(t)\right] - 
\sin{(2ht)}\text{Im}\left[L(t)\right]\right).
\eeq

\begin{figure}[htb]
\includegraphics[width=.4\textwidth]{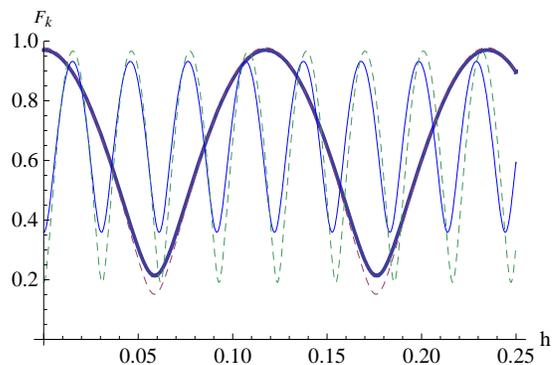}
\caption{(Color online) 
Magnetic field dependence of the fidelity for the state 
$(\ket{{\bf{0}}} + \ket{\Psi_k})/\sqrt{2}$. The solid (dashed) lines 
correspond to $k=2$ ($k=4$). The two lower (higher) frequency curves 
have been calculated for $N=51$ ($N=201$). Curves associated with 
a specific $k$ have been calculated with respect to the time 
which maximizes $\ket{\Psi_k}$. }
\label{fig:oddmag1}
\end{figure}

It is shown in the appendix that $\tilde{f}_{s,j}(t)$ 
is purely real if $s$ and $j$ are both even 
or both odd and that 
$\tilde{f}_{s,j}(t)$ is purely imaginary if $s$ is even and $j$ 
is odd or $j$ is even and $s$ is odd. This implies that 
$L(t)$ is real when $N$ is odd and that $L(t)$ is imaginary when $N$ is even. 
In either case the fidelity will oscillate with respect to $h$ with 
a frequency given by $\nu = t/\pi.$ Since $\ket{\Psi_k}$ 
reaches its maximum fidelity at roughly $t_0 \approx N/2$ 
(see Table I), the frequency of oscillation with $h$ is roughly 
$\nu \approx N/2\pi$ at the time at which $\ket{\xi_k}$ is 
to be measured. This increase in the sensitivity 
of $F_k$ to variations in $h$ as the chain grows larger can be 
seen in Fig.~\ref{fig:oddmag1}. There the solid and dashed 
lines respectively 
give the maximum fidelity for the states 
$(\ket{{\bf{0}}} + \ket{\Psi_2})/\sqrt{2}$ and 
$(\ket{{\bf{0}}} + \ket{\Psi_4})/\sqrt{2}$ as a function 
of $h$.

\begin{figure}[htb]
\includegraphics[width=.4\textwidth]{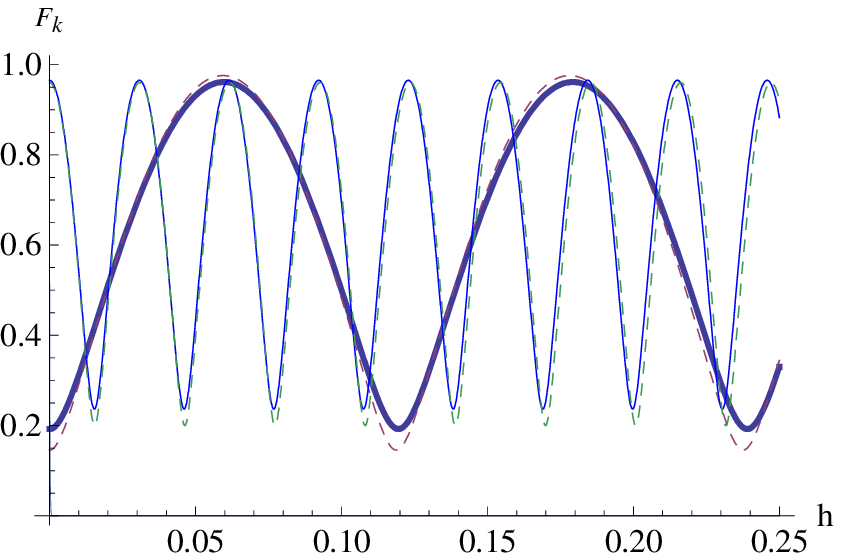}
\caption{(Color online) 
Magnetic field dependence of the fidelity for the state 
$(\ket{{\bf{0}}} + \ket{\Psi_k})/\sqrt{2}$. The solid (dashed) lines 
correspond to $k=3$ ($k=5$). The two lower (higher) frequency curves 
have been calculated for $N=51$ ($N=201$). Curves associated with 
a specific $k$ have been calculated with respect to the time 
which maximizes $\ket{\Psi_k}$. }
\label{fig:oddmag2}
\end{figure}

\begin{figure}[htb]
\includegraphics[width=.4\textwidth]{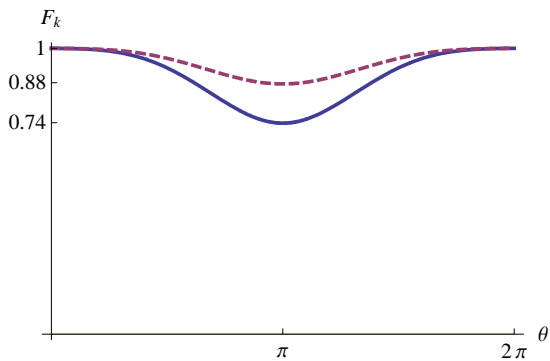}
\caption{(Color online) Fidelity when transferring the state 
$\ket{\xi_k}$ over an isolated spin chain. The solid (dashed) line 
corresponds to $k=2, N=203$ ($k=3, N=201$). Curves associated with 
a specific $k$ have been calculated with respect to the time 
which maximizes $\ket{\Psi_k}$. }
\label{fig:theta}
\end{figure}

The graph clearly shows an increase in the frequency 
as the number of spins rises from $N = 51$ to $N = 201$. Also, 
since the fidelity of the state $\ket{\Psi_k}$ is 
given by $|L(t)|$ we see a decrease in the amplitude of these 
curves for increasing $N$. (Note that the square root results in an 
asymmetry of these curves about the value corresponding to 
$\cos{(2ht)} = 0$.) The fact that the frequency grows linearly 
with $N$ suggests a technical challenge in the realization of
these
maximum fidelities for long chains. For example, for a chain 
composed of $N = 10,000$ sites a change in $h$ on the order of
$\Delta h \approx 10^{-4}$ will shift $F_k$ from its maximum 
value to its minimum value.

Instead of trying to maximize the value of $F_k$ by finely tuning some 
nonvanishing magnetic field, one could 
attempt to isolate the chain in order to achieve that same value 
if the number of sites 
are appropriately chosen. Since $L(t)$ is real when $N$ is an odd number 
the fidelity $F_k$ is either minimized or maximized when $h=0$ 
and $N$ is odd depending 
on whether $L(t)$ is positive or negative (see Eq.~({\ref{eq:fhk}})). For 
a given $k$ the sign of $L(t)$ alternates as the number of spins 
is increased by two. If $k$ is even the fidelity will be maximized 
(i.e, $L(t)$ is strictly positive) when $h=0$ if $N$ can be written 
as $N= 3+4m$ for some integer $m$, otherwise $F_k$ will be minimized at $h=0$. 
Similarly, if $k$ is odd the fidelity will be maximized 
when $h=0$ if $N$ can be written as $N= 1+4m^{\prime}$ 
for some other integer $m^{\prime}$, otherwise $F_k$ will 
be minimized at $h=0$

Figure 3 shows the magnetic field dependence of the fidelity 
for the states $(\ket{{\bf{0}}} + \ket{\Psi_3})/\sqrt{2}$ 
and $(\ket{{\bf{0}}} + \ket{\Psi_5})/\sqrt{2}$. 
The two higher frequency curves are nearly 
indistinguishable since 
the fidelities for $\ket{\Psi_3}$ and $\ket{\Psi_5}$,
and thus the two amplitudes, 
are nearly equal for $N=201$ (see Table~I).

When the applied field $h$ is chosen such that it maximizes 
$F_k$ the fidelity of the state $\ket{\xi_k}$ decreases 
smoothly from unity when $\theta=0$ to a minimum value 
when $\ket{\xi_k} = \ket{\Psi_k}$. (Notice that $F_k$ is 
independent of $\phi$.) This behavior is illustrated in 
Fig.~\ref{fig:theta} for $k=2$ and $k=3$. The fidelities shown in that 
figure 
have been calculated for isolated chains ($h=0$) 
containing $N=203$ and $N=201$ respective sites.

\begin{figure}[htb]
\includegraphics[width=.4\textwidth]{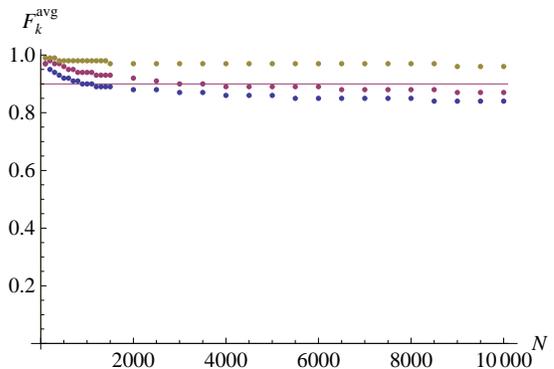}
\caption{(Color online) 
Average fidelity of the states $\ket{\xi_k}$ 
for k=2,3, and 11. (From bottom to top.  There are no crossings.)  
The horizontal line lies at 90$\%$. } 
\label{fig:Favg}
\end{figure}

Since $\ket{{\bf{0}}}$ is an eigenstate of the Hamiltonian 
the average fidelity of $\ket{\xi_k}$ will be greater 
than the fidelity of $\ket{\Psi_k}$ alone when $h$ is chosen 
to maximize $F_{h,k}$. This occurs when $h$ takes on a value such 
that the product $\cos{(2ht)}{\mbox{Re}}[L(t)]$ or 
$\left(-\sin{(2ht)}{\mbox{Im}}[L(t)]\right)$ 
is positive and equal to $|L(t)|$. At these points the fidelity $F_k$ 
becomes
\bea
\label{eq:Fmax}
F_{k}^{max} &=& \left[ \cos^4{\frac{\theta}{2}} 
+  \frac{\sin^2{\theta}}{2}\left(
|L(t)| + \frac{1}{k} \sum_{m=1}^{N+1-2k}|C_{m}(t)|^2 \right) \right. 
\nonumber \\ 
&+& \left. \sin^4{\frac{\theta}{2}}|L(t)|^2 \right]^{1/2}.
\eea  
The average fidelity of $F_{k}^{max}$ can be calculated as 
\beq
\label{eq:Favg}
F_{k}^{avg} = \frac{1}{2}\int \limits_{0}^{\pi} F_{k}^{max} 
\sin{(\theta)} d \theta
\eeq
Figure~\ref{fig:Favg} exemplifies the high average fidelities which
can be obtained when transferring the states $\ket{\xi_k}$ 
over long chains. Even for chains containing 10,000 sites the 
lowest value of the average fidelity, corresponding to 
$\ket{\xi_2}$, is still roughly $85\%$. For an $N=10,000$ spin chain 
the average fidelities associated with the states $\ket{\xi_3}$ and 
$\ket{\xi_{11}}$ are respectively $87\%$ and $96\%$. 

%-----------------------------------------------------------
%-----------------------------------------------------------

\section{Optimal State Encoding}

We will now compare our results to those which can be obtained 
using Haselgrove's optimal encoding scheme \cite{Haselgrove:05}. 
Specifically, we will compare the maximum fidelities 
that can be obtained in either case for initial encodings 
that take place over equal regions of the chain.
 
In what follows we 
will assume that Alice and Bob respectively have access to the first 
and last $r$ sites of a linear spin chain and that the system 
evolves according to the Hamiltonian 
given by Eq.~(\ref{eq:Ham}). Using the notation of 
\cite{Osborne/etal:04,Haselgrove:05} 
we may express an arbitrary initial encoding of the chain as 
\beq
\ket{\Psi(0)} = (\alpha \ket{{\bf{0}}}_A +\beta \ket{1_{enc}}_A)
\otimes \ket{{\bf{0}}}_{\bar{A}},
\eeq
where $\ket{1_{enc}}_A \: \in \: \:{\cal{H}}^{(1)}$ 
and $A$ ($\bar{A}$) refers to the spins which Alice does (does 
not) control. Since $\ket{{\bf{0}}}$ is an eigenstate of the 
system we may write the evolved state 
$\ket{\Psi(T)} = e^{-iHT}\ket{\Psi(0)}$ in the general form 
\bea
\ket{\Psi(T)} &=& \beta \sqrt{1-C_B(T)}\ket{\eta(T)}  \\
&+& \ket{{\bf{0}}}_{\bar{B}}[\alpha \ket{{\bf{0}}}_B + \beta 
\sqrt{C_B(T)}\ket{\gamma(T)}_B].\nonumber
\eea 
Here $C_B(T)$ gives the probability that the excitation has 
transferred into the receivers accessible region. 
$\ket{\gamma(T)}_B$ is a some normalized state that is 
orthogonal to $\ket{{\bf{0}}}_B$ and $\ket{\eta(T)}$ is some 
normalized state that is orthogonal to all states of the form 
$\ket{{\bf{0}}}_{\bar{B}} \otimes \ket{\nu}_B$.  As was mentioned 
in the introduction, according to this protocol the receiver 
will apply a decoding unitary to his accessible spins such that the 
probability $C_B(T)$ will be transferred to a single spin site. 
After this has occurred the fidelity between the final output 
qubit state and the intended message state can be expressed as 
\beq
\left[|\alpha|^4 +[1+2\sqrt{C_B(T)}-C_B(T)]|\alpha|^2 |\beta|^2 + 
C_B(T)|\beta|^4\right]^{1/2}
\eeq
(We note that the global square root does not appear in 
Ref.~\cite{Osborne/etal:04}.) 
Since we are only concerned with the $\ket{1_{enc}}_A$ states, 
we will set $\alpha = 0$ and $\beta = 1$. The fidelity 
associated with an encoding $\ket{1_{enc}}_A$ over the 
first $r$ spins of the chain is then 
\beq
F_r = \sqrt{C_B(T)}.
\eeq

If we let $P_{\cal{A}}$ and $P_{\cal{B}}$ represent the projectors 
onto Alice's and Bob's accessible subspaces we may express 
$C_B(T)$ as 
\beq
C_B(T) = ||P_{\cal{B}}\:e^{-iHt}\ket{1_{enc}}_A
\otimes \ket{{\bf{0}}}_{\bar{A}}||^2,
\eeq
where $||\cdot||$ represents the $l_2$ norm. Haselgrove's optimal 
states are those which maximize $C_B(T)$. They correspond 
to the first right singular vectors of $P_{\cal{B}}\:e^{-iHt}\:P_{\cal{A}}$.
For an encoding which uses $r$ spins, we will refer to the optimal 
state as $\ket{\Phi}_r$. 

Table~\ref{tab:f2} provides the maximum 
fidelities which can be obtained for the states 
$\ket{\Phi}_r$ ($r=3,5,7,9$) as a function of $N$. The difference 
between these maximum values and those associated with the 
states $\ket{\Psi}_{(r+1)/2}$ is given in terms of $\Delta_r$. We 
see that as the ratio of the number of encoded spins to 
the total number of spins approaches zero the value of $\Delta_r$ 
also goes to zero. The table also gives 
a measure of how close these states resemble the $\ket{\Psi}_k$ in 
terms of the quantity $d_r=|\ket{\Phi}_r - \ket{\Psi}_{(r+1)/2}|$. For 
large chains our states actually converge to the optimal states 
obtained using Haselgrove's method. It is interesting to note that 
in all cases the times which maximize the fidelity is nearly the 
same for both our states and Haselgrove's states for encodings that
take place over the first $r$ spins. The fact that $\Delta_r$ approaches 
zero for large chains is also interesting since the fidelities which 
were calculated for our states were taken place before any decoding 
operation while the fidelities associated with Haselgrove's states 
were determined after this assumed operation had occurred. This 
implies that for large chains the fidelities 
associated with these optimal states are independent of the decoding 
operation.

\begin{table}[tbp]

\begin{tabular}{ccccccccccccccccccccccccccccccc}   
 && && &&  $\ket{\Phi}_3$  && && &&

\end{tabular} 

\begin{tabular}{|c|cccccccccccc|cccccccccccccccccc}

\hline 
$N$    && 100  && 200  && 300   && 400   && 500   && 3000  

\\ \hline
$F_3$    && .85  && .75 && .68  && .64  && .60  &&   .36

\\ \hline
$\Delta_3$    && .02  && .01  && .00  && .00  && .00  &&  .00

\\ \hline
$d_3$     && .13  && .09  && .07  && .06  && .05   && .02

\\ \hline
$t_0$ && 51.75 && 102.36 && 152.76 && 203.07 && 253.33 &&  1506.14    \\

\hline

\end{tabular} 

\begin{tabular}{ccccccc}   \\

 $\ket{\Phi}_5$  \\  

\end{tabular} 

\begin{tabular}{|c|cccccccccccc|cccccccccccccccccc}  \hline

$N$    && 100  && 200  && 300   && 400   && 500   && 3000

\\ \hline
$F_5$    && .96  && .91  && .86  && .82  && .79  && .52

\\ \hline
$\Delta_5$    && .06  && .03  && .02  && .01  && .01  && .01

\\ \hline
$d_5$     && .27 && .19 && .15  && .13  && .11   && .04

\\ \hline
$t_0$ && 51.08 && 101.74 && 152.20 && 202.55 && 252.84 &&   1505.83

\\ \hline 

\end{tabular} 

\begin{tabular}{ccccccccccccc|cccccccccccccccccc} \\

$\ket{\Phi}_7$ 

\end{tabular} 

\begin{tabular}{|c|cccccccccccc|cccccccccccccccccc}  \hline

$N$    && 100  && 200  && 300   && 400   && 500   && 3000   

\\ \hline
$F_7$    &&.99  && .97  && .95  && .92  && .90  &&  .64

\\ \hline
$\Delta_7$    && .11   && .07  && .05  && .04  && .03  &&  .01

\\ \hline
$d_7$     && .40 && .29 && .25  && .21  && .19   && .07

\\ \hline
$t_0$ && 50.25 && 101.00 && 151.45 && 201.89 && 252.20 && 1505.51

\\ \hline

\end{tabular} 

\begin{tabular}{ccccccccccccc|cccccccccccccccccc}  \\

$\ket{\Phi}_9$  

\end{tabular} 

\begin{tabular}{|c|cccccccccccc|cccccccccccccccccc}  \hline

$N$    && 100  && 200  && 300   && 400   && 500   && 3000   

\\ \hline
$F_9$    && .99 && .99  && .98  && .97  && .96  && .75

\\ \hline
$\Delta_9$    && .06   && .11  && .10  && .07  && .06  &&  .01

\\ \hline

$d_9$     && .52  && .40  && .34  && .29   && .26  &&  .12

\\ \hline
$t_0$ && 49.4 && 100.21 && 150.74 && 201.14 && 251.49 && 1505.00

\\ \hline 
\end{tabular}

\caption{Table showing the maximum fidelity achievable and the associated 
arrival times for the states $\ket{\Phi_r}$, $r=3,5,7,9$ as 
a function of $N$. $\Delta_r$ gives the difference between 
the maximum fidelities associated with 
$\ket{\Phi_r}$ and $\ket{\Psi_(r+1)/2}$. The quantity 
$d_r=|\ket{\Phi}_r - \ket{\Psi}_{(r+1)/2}|$ is also given. 
Note that the time has been 
calculated using $J=1.$ }
\label{tab:f2}
\end{table}

%-----------------------------------------------------------
%-----------------------------------------------------------

\section{conclusion}

The $k$-qubit encodings which we have introduced can be used 
to reliably transmit information over very large spin chains. Their 
simple structure was shown to yield high transfer fidelities 
after the system was allowed to evolve freely for some specified time. 
These successful transmissions did not require the dynamical 
control or individual design of the exchange couplings, nor did 
they require the implementation of decoding operations at the 
receiving end. When the chains are placed in a global magnetic field 
the fidelities associated with these states were found to become highly 
sensitive to fluctuations in the field strengths as the chains 
become large. The frequency of the fidelities oscillation with 
the field was determined to be proportional 
to the number of spins in the chain suggesting the 
technical difficulties associated with achieving these high values 
for large $N$. It was found that 
if the chain contained an appropriate number of sites the 
fidelity could be maximized when the chain was isolated from the 
field. When $k$ is an even number the fidelity will 
take its maximum value  
when $h=0$ if $N$ can be written as $N= 3+4m$ for some integer $m$. 
Similarly, if $k$ is odd the fidelity will be maximized 
when $h=0$ if $N$ can be written as $N= 1+4m^{\prime}$ 
for some other integer $m^{\prime}$.   
 
A comparison with Haselgrove's optimal encodings \cite{Haselgrove:05} 
was also given. It was shown that for small chains his 
states will yield slightly higher fidelities than ours when the 
encodings are viewed to take place over the first $r$ spins of the chain. 
However, since our states only require the initialization of $k$ of the 
first $r =2k - 1$ spins, preparation of our states would appear 
easier for near-future realization. For large chains our states 
converge to Haselgrove's optimal encodings, and can transfer with  
fidelities that are independent of a decoding stage. It is interesting 
that these even site encodings yield much higher fidelities when 
compared to the analogous odd site encodings. However, the fundamental reason 
for their success in propagation remains unclear.

Given the simple structure inherent to our encodings, along with 
the minimal technical requirements needed for reliable transmission, 
we believe that these states could serve as useful message carriers 
over large spin chains.

%-----------------------------------------------------------
%-----------------------------------------------------------

\section{Appendix}

We will show here that the $\tilde{f}_{s,j}(t)= 
\frac{2}{N+1}\sum_{m=1}^N \sin{(q_m s)} 
\sin{(q_m j)} e^{2iJt\cos{(q_m)}}$ are either purely real 
or purely imaginary. First notice that for $s=1,2,...,N$ 
and $\delta = 0,1,2,...,N-1$ we have
\bea
\label{eq:a1}
\sin{(q_{(N - \delta)}s)} &=& \sin{(q_s(N-\delta))}
\nonumber \\
&=& \sin{(q_s N)}\cos{(q_s \delta)} - \cos{(q_s N)}\sin{(q_s \delta)},
\nonumber \\
\eea
with
%$
%\sin{(q_s N)} = \sin{(\pi s)}\cos{(q_s)} - 
%\cos{(\pi s)}\sin{(q_s)} \nonumber \\
%= (-1)^{(s+1)}\sin{(q_s)},
%$
$$\sin{(q_s N)} = (-1)^{s+1}\sin{(q_s)}$$
and 
%\bea
%\cos{(q_s N)} &=& \cos{(\pi s)}\cos{(q_s)} + \sin{(\pi s)}\sin{(q_s)}
%\nonumber \\
%&=& (-1)^s \cos{(q_s)}.
%\eea
$$\cos{(q_s N)} = (-1)^s \cos{(q_s)}.$$
%So we may write Eq.~(\ref{eq:a1}) as 
%\bea
%\sin{(q_{(N - \delta)}s)} &=& (-1)^{s+1}[\sin{(q_s)}\cos{(q_s \delta)} 
%\nonumber \\ \;\;\;
%&+& \cos{(q_s)}\sin{(q_s \delta)}] \nonumber \\
%&=& (-1)^{s+1}\sin{(q_{(1 + \delta)}s)}
%\eea
So we have the identity
\beq
\label{eq:a2}
\sin{(q_{(N - \delta)}s)} = (-1)^{s+1}\sin{(q_{(1 + \delta)}s)}.
\eeq
It can also be shown that the following relation holds
\beq
\label{eq:a3}
\cos{(q_{(N - \delta)})} = -\cos{(q_{(1 + \delta)})}.
\eeq
Equation~(\ref{eq:a2}) implies that  
$\sin{(q_{(1+\delta)}s)} \sin{(q_{(1+\delta)}j)} =  
\sin{(q_{(N - \delta)}s)} \sin{(q_{(N - \delta)}j)}$ if $s$ and $j$ 
are both even or both odd, and that 
$\sin{(q_{(1+\delta)}s)} \sin{(q_{(1+\delta)}j)} =  
-\sin{(q_{(N - \delta)}s)} \sin{(q_{(N - \delta)}j)}$ if $s$ 
or $j$ is even while the other is odd. Let us now consider four cases. 

Case (1):  The number of spins $N$ is even 
while $s$ and $j$ are both even or both odd. In this case 
the $(1+ \delta) th$ term 
of the series $\text{Re}\left[\tilde{f}_{s,j}(t)\right]= 
\frac{2}{N+1}\sum_{m=1}^N \sin{(q_m s)} 
\sin{(q_m j)} \cos{(2Jt\cos{(q_m)})}$ will equal the 
$(N - \delta) th$ term since $\cos{(-x)} = \cos{(x)}$. Also, 
since $N$ is even every term in the series can be matched with 
another term. Since $\sin{(x)} = -\sin{(-x)}$ the $(1+ \delta) th$ term 
of the series $\text{Im}\left[\tilde{f}_{s,j}(t)\right]= 
\frac{2}{N+1}\sum_{m=1}^N \sin{(q_m s)} 
\sin{(q_m j)} \sin{(2Jt\cos{(q_m)})}$ will cancel with the 
$(N - \delta) th$ term. $N$ is even so every term will cancel. In this 
case $\tilde{f}_{s,j}(t) = \text{Re}\left[\tilde{f}_{s,j}(t)\right]$.

Case (2):  The number of spins $N$ is odd 
while $s$ and $j$ are both even or both odd. The situation here 
is the same as for the previous case except now one has to account for 
the $(N+1)/2 \:th$ term. Since $\cos{(2Jt\cos{(q_{(N+1)/2})})} = 1$ 
and $\sin{(2Jt\cos{(q_{(N+1)/2})})} = 0$ we find that 
$\text{Im}\left[\tilde{f}_{s,j}(t)\right]=0$ again 
for this second case. Consequently  
$\tilde{f}_{s,j}(t) = \text{Re}\left[\tilde{f}_{s,j}(t)\right]$.

Case (3):  $N$ and $s$ are even 
while $j$ is odd or $N$ and $j$ are even while $s$ is odd. In 
this case the $(1+ \delta) th$ term will equal (cancel) the 
$(N - \delta) th$ term of the series expansion for
$\text{Im}\left[\tilde{f}_{s,j}(t)\right]$ 
($\text{Re}\left[\tilde{f}_{s,j}(t)\right]$). $N$ is assumed 
to be even here so every term in $\text{Re}\left[\tilde{f}_{s,j}(t)\right]$ 
will cancel with another term. In this case
$\tilde{f}_{s,j}(t) = \text{Im}\left[\tilde{f}_{s,j}(t)\right]$.

Case (4):  $N$ and $s$ are odd 
while $j$ is even or $N$ and $j$ are odd while $s$ is even. Again, the 
only difference between this case and the last is due to the fact that the 
$(N+1)/2 \:th$ term cannot be matched up with another term. This middle 
term is zero for $\text{Im}\left[\tilde{f}_{s,j}(t)\right]$ as it was 
in case two. For $\text{Re}\left[\tilde{f}_{s,j}(t)\right]$ this term 
is equal to $\frac{2}{N+1}\sin{(\pi s/2)}\sin{(\pi j/2)} = 0$ since either 
$s$ or $j$ is even. So for this case 
 $\tilde{f}_{s,j}(t) = \text{Im}\left[\tilde{f}_{s,j}(t)\right]$.

In every case $\tilde{f}_{s,j}(t)$ is either purely real or 
purely imaginary. Whether $\tilde{f}_{s,j}(t)$ is real or imaginary 
depends on the indices $s$ and $j$. Regardless of the number of 
spins in the chain, $\tilde{f}_{s,j}(t)$ is real if $s$ and $j$ are both even 
or both odd and $\tilde{f}_{s,j}(t)$ is imaginary if $s$ is even and $j$ 
is odd or $j$ is even and $s$ is odd.

%-----------------------------------------------------------
%-----------------------------------------------------------

%\bibliography{bib2}

\begin{thebibliography}{11}
\expandafter\ifx\csname natexlab\endcsname\relax\def\natexlab#1{#1}\fi
\expandafter\ifx\csname bibnamefont\endcsname\relax
  \def\bibnamefont#1{#1}\fi
\expandafter\ifx\csname bibfnamefont\endcsname\relax
  \def\bibfnamefont#1{#1}\fi
\expandafter\ifx\csname citenamefont\endcsname\relax
  \def\citenamefont#1{#1}\fi
\expandafter\ifx\csname url\endcsname\relax
  \def\url#1{\texttt{#1}}\fi
\expandafter\ifx\csname urlprefix\endcsname\relax\def\urlprefix{URL }\fi
\providecommand{\bibinfo}[2]{#2}
\providecommand{\eprint}[2][]{\url{#2}}

\bibitem[{\citenamefont{{S. Bose}}(2003)}]{Bose:03}
\bibinfo{author}{\bibnamefont{{S. Bose}}}, \bibinfo{journal}{Phys. Rev. Lett.}
  \textbf{\bibinfo{volume}{91}}, \bibinfo{pages}{{207901}}
  (\bibinfo{year}{2003}).

\bibitem[{\citenamefont{{L.-A. Wu, Y. Liu, F. Nori}}(2009)}]{Wu/etal:09}
\bibinfo{author}{\bibnamefont{{L.-A. Wu, Y. Liu, F. Nori}}}
  (\bibinfo{year}{2009}), \bibinfo{note}{{arXiv:quant-ph/0903.2154}}.

\bibitem[{\citenamefont{{G. M. Nikolopoulos, D. Petrosyan and P.
  Lambropoulos}}(2004)}]{Petrosyan:04a}
\bibinfo{author}{\bibnamefont{{G. M. Nikolopoulos, D. Petrosyan and P.
  Lambropoulos}}}, \bibinfo{journal}{Europhys. Lett.}
  \textbf{\bibinfo{volume}{65}}, \bibinfo{pages}{297} (\bibinfo{year}{2004}).

\bibitem[{\citenamefont{{G.M. Nikolopoulos, D. Petrosyan, P.
  Lambropoulos}}(2004)}]{Petrosyan:04b}
\bibinfo{author}{\bibnamefont{{G.M. Nikolopoulos, D. Petrosyan, P.
  Lambropoulos}}}, \bibinfo{journal}{{J. Phys: Cond. Mat.}}
  \textbf{\bibinfo{volume}{16}}, \bibinfo{pages}{4991} (\bibinfo{year}{2004}).

\bibitem[{\citenamefont{{M. Christandl, N. Datta, A. Ekert, and A.J.
  Landahl}}(2004)}]{Christandl/etal:04}
\bibinfo{author}{\bibnamefont{{M. Christandl, N. Datta, A. Ekert, and A.J.
  Landahl}}}, \bibinfo{journal}{Phys. Rev. Lett.}
  \textbf{\bibinfo{volume}{92}}, \bibinfo{pages}{{187902}}
  (\bibinfo{year}{2004}).

\bibitem[{\citenamefont{{M. Christandl, N. Datta, T.C. Dorlas, A. Ekert, A.
  Kay, and A.J. Landahl}}(2005)}]{Christandl/etal:05}
\bibinfo{author}{\bibnamefont{{M. Christandl, N. Datta, T.C. Dorlas, A. Ekert,
  A. Kay, and A.J. Landahl}}}, \bibinfo{journal}{Phys. Rev. A}
  \textbf{\bibinfo{volume}{71}}, \bibinfo{pages}{{032312}}
  (\bibinfo{year}{2005}).

\bibitem[{\citenamefont{{H.L. Haselgrove}}(2005)}]{Haselgrove:05}
\bibinfo{author}{\bibnamefont{{H.L. Haselgrove}}}, \bibinfo{journal}{Phys. Rev.
  A} \textbf{\bibinfo{volume}{72}}, \bibinfo{pages}{{062326}}
  (\bibinfo{year}{2005}).

\bibitem[{\citenamefont{{D. Burgarth, V. Giovannetti, and S.
  Bose}}(2007)}]{Burgarth:07}
\bibinfo{author}{\bibnamefont{{D. Burgarth, V. Giovannetti, and S. Bose}}},
  \bibinfo{journal}{Phys. Rev. A} \textbf{\bibinfo{volume}{75}},
  \bibinfo{pages}{{062327}} (\bibinfo{year}{2007}).

\bibitem[{\citenamefont{{C. Di Franco, M. Paternostro, and M.S.Kim}}(2008)}]
{Franco/etal:08}
\bibinfo{author}{\bibnamefont{{C. Di Franco, M. Paternostro, and M.S.Kim}}},
  \bibinfo{journal}{Phys. Rev. Lett.} \textbf{\bibinfo{volume}{101}},
  \bibinfo{pages}{{230502}} (\bibinfo{year}{2008}).


\bibitem[{\citenamefont{{T.J. Osborne and N. Linden}}(2004)}]{Osborne/etal:04}
\bibinfo{author}{\bibnamefont{{T.J. Osborne and N. Linden}}},
  \bibinfo{journal}{Phys. Rev. A} \textbf{\bibinfo{volume}{69}},
  \bibinfo{pages}{{052315}} (\bibinfo{year}{2004}).

\bibitem[{\citenamefont{{J. Allcock and N. Linden}}(2009)}]{Allcock/etal:09}
\bibinfo{author}{\bibnamefont{{J. Allcock and N. Linden}}},
  \bibinfo{journal}{Phys. Rev. Lett.} \textbf{\bibinfo{volume}{102}},
  \bibinfo{pages}{{110501}} (\bibinfo{year}{2009}).

\bibitem[{\citenamefont{Wang et~al.}(2009)\citenamefont{Wang, Bishop, Byrd,
  Shao, and Zou}}]{Wang/etal:09}
\bibinfo{author}{\bibfnamefont{Z.-M.} \bibnamefont{Wang}},
  \bibinfo{author}{\bibfnamefont{C.~A.} \bibnamefont{Bishop}},
  \bibinfo{author}{\bibfnamefont{M.~S.} \bibnamefont{Byrd}},
  \bibinfo{author}{\bibfnamefont{B.}~\bibnamefont{Shao}}, \bibnamefont{and}
  \bibinfo{author}{\bibfnamefont{J.}~\bibnamefont{Zou}},
  \bibinfo{journal}{Phys. Rev. A} \textbf{\bibinfo{volume}{80}},
  \bibinfo{eid}{022330} (\bibinfo{year}{2009}).

\bibitem[{\citenamefont{{S. Bose}}(2007)}]{Bose:07}
\bibinfo{author}{\bibnamefont{{S. Bose}}},
  \bibinfo{journal}{Contemp. Phys.} \textbf{\bibinfo{volume}{48}},
  \bibinfo{pages}{{13}} (\bibinfo{year}{2007}).

\bibitem[{\citenamefont{{M. Hase, H. Kuroe, K. Ozawa, O. Suzuki, H. Kitazawa, G. Kido, and T. Sekine}}(2004)}]{Hase/etal:04}
\bibinfo{author}{\bibnamefont{{M. Hase, H. Kuroe, K. Ozawa, O. Suzuki, H. Kitazawa, G. Kido, and T. Sekine}}},
  \bibinfo{journal}{Phys. Rev. B} \textbf{\bibinfo{volume}{70}},
  \bibinfo{pages}{{104426}} (\bibinfo{year}{2004}).
\end{thebibliography}

%-------------------------------------------------------------------------
%-------------------------------------------------------------------------

\end{document}